\documentclass{cernphprep}
\usepackage{cite,fancyvrb}
\usepackage{cernunits}
\usepackage{graphicx}
\usepackage{dcolumn}
\usepackage{bm}
\usepackage{color}

\newcommand\ee{\end{equation}}
\newcommand\be{\begin{equation}}
\newcommand\eea{\end{eqnarray}}
\newcommand\bea{\begin{eqnarray}}

\begin{document}
\begin{titlepage}
\PHdate{\today}

\title{Kinematic Analysis Towards Glueballs}

\vspace{1mm}

\begin{Authlist}
Simone Giani\Aref{a}, Hubert Niewiadomski\Aref{b}
\Instfoot{a1}{CERN, Geneva, Switzerland}
Luca Trentadue\Aref{c}
\Instfoot{a2}{Dipartimento di Fisica e Scienze della Terra ``M. Melloni'' \\Universit\'a di Parma, Parma, Italy\\
and\\
INFN, Sezione di Milano Bicocca, Milano, Italy}
\end{Authlist}

\vspace{10mm}

\begin{abstract}
In the present work  a consistent kinematic-based framework for glueball states is proposed. It relates
the glueball, the Pomeron, QCD lattice calculations, the $0^{++}$ scalar states $f_0(1710)$ and $\chi_{c0}(1P)$, the $2^{++}$ states $f_J(2220)$ and $\chi_{c2}(2P)$, the baryonic charmed state $\Xi_c^+(2645)$.
\end{abstract}

\Anotfoot{a}{simone.giani@cern.ch}
\Anotfoot{b}{hubert.niewiadomski@cern.ch}
\Anotfoot{c}{luca.trentadue@cern.ch}
\end{titlepage}


\section{Theoretical and phenomenological framework}
Glueballs are predicted by Quantum Chromodynamics (QCD)   as gluon bound states with no quark contents~\cite{GELLM,MATHI,YCHEN,SEEPL}. QCD lattice calculations foresee a $J^{PC}= 0^{++}$ ground state and a $2^{++}$ state followed by a spectrum of excited states as well as exotic combinations~\cite{YCHEN,MORNI,MATHI,MEYER,WTOKI,WOCHS}. Glueballs were also predicted as states on the extrapolation to positive values of the Mandelstam variable $t$ of the Pomeron Regge trajectories, accounting for the fact that Pomerons were never observed in final states~\cite{HOTHI,BICUD,ZHOUL}.

Searches for glueballs were reported already at the CERN ISR~\cite{ABREA} and at fixed target experiments~\cite{SINGO,AMSLE,AUSTR,HASAL} by doing spectroscopic analyses. Different production channels, so-called "gluon-rich"~\cite{WEIDA,ZHENG,BERNH,MURGI,WOCHS}, have been chosen~\cite{MATHI,THOMA,ALBRO,WTOKI,CREDE}: $J/\psi$ radiative processes~\cite{PDG12,FRANK}, $p\bar{p} $  annihilation~\cite{PDG12,AMSLE}, Double Pomeron Exchange (DPE)~\cite{ROBSO,ABREA,KIPLB}, as well as semi-leptonic channels~\cite{PDG12}. An extended spectrum of scalar/tensorial states have been measured in the  $1-2\UGeV$  range by several experiments~\cite{PDG12,THOMA,WEIDA,WTOKI,ABREA,WA102,FRANK,AMSLE,ZHENG,MACHA}: 
$f_0(980), f_2(1270), f_0 (1370), f_0(1500), f_0(1710), f_2(1910), ... $ up to the $f_2(2340)$. 
These states have $0^{++}$ or $2^{++}$  $J^{PC}$  quantum numbers and may constitute an excess as meson states with respect to the $SU(3)_f$ quark-based nonet~\cite{SEEPJ,ZHENG,SINGO,CREDE}. Therefore they have been considered and analysed as glueball candidates~\cite{WEIDA,HAIYA,KINPA,ZHENG,CREDE} based on their decay modes and branching ratios, including also the possibility of different mixing scenarios between the glueball mass states and the nearby mesonic states~\cite{MATHI,HAIYA,DEWIT,KAIDA,ZHENG,STROH,CREDE,HASHI}. Several works~\cite{BICUD,MEYER,MATHI,KAIDA,DOLAN,BOSCH,SEEPJ,SEEPL,HOTHI,ZHOUL,ESTRA} have also analysed consistent parameters of Pomeron's Regge leading and subleading trajectories with those, respectively, of the $2^{++}$ and $0^{++}$ glueball candidates. 

A growing stream of theoretical interest is looking beyond the apparently coincidental consistency of the Pomeron and glueballs Regge trajectories~\cite{BICUD,KAIDA,BJORK,BOSCH,SEEPJ,SEEPL,HOTHI,ZHOUL}. Under certain conditions, the Pomeron can be considered as a pure combination (pair, ladder,\dots) of color-compensating gluons~\cite{BJORK,MATHI,DOLAN}, with no quark nor Reggeon contamination. In these cases, the relation between the Pomeron and the two-gluons glueball~\cite{MATHI,BICUD,HOTHI,ZHOUL} can be formalized/described via a duality between the $s$-channel and the $t$-channel as if the Pomeron were a virtual glueball or the glueball were a real Pomeron~\cite{SEEPJ,SEEPL,BICUD,HOTHI,ZHOUL,BJORK}. 

Consequently a kinematic analysis of the glueballs' production channels and of their eventual related enhancements can become complementary to the decay and related suppression criteria 
(for which, although starting from flavor symmetric amplitudes, the phase space and the mass coupling induce opposite trends)~\cite{DEWIT,THOMA,WEIDA,JCAOT,CREDE,HASHI,CHANO} as discriminant factors in the determination of glueball-candidates.

Furthermore it can be noted that the Pomeron, being colorless, is a composite state of gluons and represents an effective mediator of the long-range~\cite{TRENT} interactions between hadronic states~\cite{ZHOUL,HOTHI}. The binding energy between colorless states, if considered as a residual of the strong force likewise is the case for nucleons in nuclei, can be neglected, being of the order  of $\sim1\%$ with respect to the masses of the interacting states. 

The present work is aimed at showing that a pattern of different independent hypotheses regarding glueballs are actually consistently supporting each other within the same framework. The framework is based only on kinematic constraints, the properties of the strong interaction dynamics discussed above, and the $s$ -- $t$ duality of the description of the interactions. 

\section{Kinematics and resonances}
\begin{figure}
\centering
\includegraphics[scale=0.4]{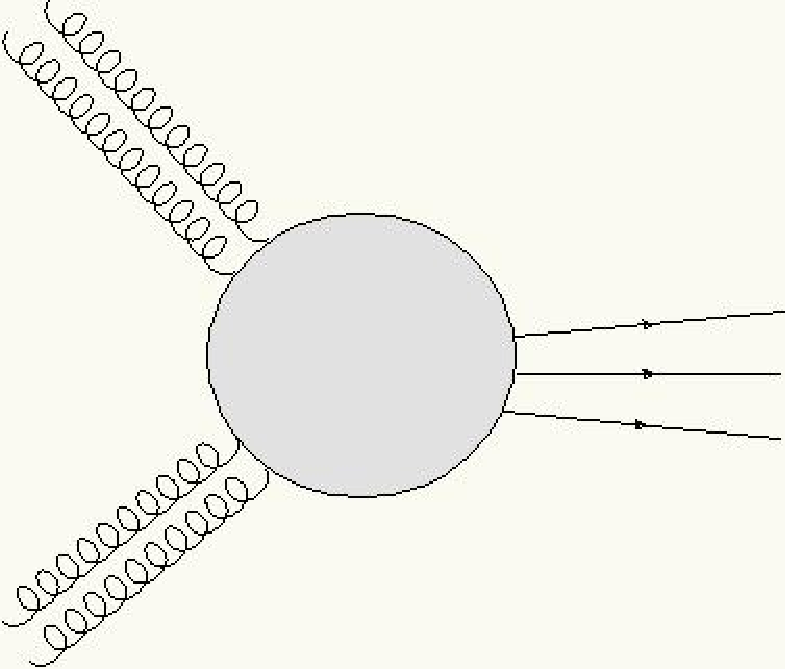}
\caption{\label{fig:epsart1}  A symbolic representation of Pomeron-Pomeron collision in either $s$ or $t$ channels. 
}
\label{dpe}
\end{figure}

The Double Pomeron Exchange (DPE) is an ideal channel to study selectively Pomeron-Pomeron scattering and in particular exclusive gluonic interactions. Figure~\ref{dpe} shows a typical DPE diagram where two protons produce a diffractive exclusive central system.

With the Pomerons treated as general virtual states, DPE can thus produce via Pomeron fusion any real $0^{++}$ mass state allowed by quantum numbers likewise the $2^{++}$ one, including a single glueball. Values of the mass of the produced diffractive system $M_x \cong \sqrt{ s\;\xi_1\;\xi_2}$  can be attained via any combination of the fractional momenta $\xi_{1,2}$ which represent $(1 - x_{1,2F})$, with $x_{1,2F}$ the Feynman $x_F$ variable of the interacting protons, where $\sqrt{s}$ is their center of mass scattering energy.

The DPE cross-section is described by the following expression~\cite{BUCHL}:

\begin{equation}\label{dde}
\sigma_{\cal{DPE}}(s)\;=\;\int\;F_1^{\cal{P}}(\xi_1,\hat{s})\;F_2^{{\cal{P}}}(\xi_2,\hat{s})\;\;\hat{\sigma}_{\cal{P}\cal{P}}(\hat{s},\xi_1,\xi_2)  \;d\xi_1\;d\xi_2   
\newline   
\end{equation}


with $F_{1,2}^{\cal{P}}(\xi_{1,2},\hat{s})$ being the diffractive structure functions of the protons, $\hat{s}$ being the reduced center-of-mass energy squared of the interacting Pomerons, and $\hat{\sigma}_{\cal{P}\cal{P}}(\hat{s},\xi_1,\xi_2)$ being the Pomeron fusion cross-section.

If the Pomerons are to be considered as virtual glueballs (i.e. the glueballs are to be considered as the real counterpart of the Pomerons)
\cite{BICUD,BJORK,SEEPJ,ZHOUL}, the consequence is that they also exist as states taking part, as glueballs, to the dual $s$-channel processes as in a Gedanken experiment of glueball-glueball collisions. 
In particular this includes the case they are on the mass-shell, 
i.e. $P_{1,2}^2 \cong M_{gg}^2$ with $M_{gg}$ being the glueball mass.  In this case, the total four-momentum carried by the real
Pomerons, induces the production of physics states respecting the condition given by the relation:
\begin{equation}\label{2}
(P_1 + P_2 )^2 \geq (2 M_{gg})^2,
\end{equation}

where $P_{1,2}$  are the four-momenta of the two real Pomerons and  $M_{gg}$ is the glueball mass. It may be noted here that a single glueball cannot be produced exclusively in these conditions given that the four-momenta obey to the constraint that both real Pomerons are close to the glueball mass shell. Most importantly, it may be observed that the relation~(\ref{2}) allows a production threshold for a $0^{++}$ physics state at twice the glueball mass, when the scalar-product of the real Pomerons' four-momenta reaches its minimal value\footnote{It is worth observing that a minimal four-momenta scalar product, implying a maximal three-momenta scalar product, means that the vector-difference of the real Pomerons' three-momenta is minimal: this may be related to the empirical correlation noted between resonances' gluonic content and real Pomerons' transverse momenta vector-difference in Refs.~\cite{KINPA,KIPLB,CLOSE,BARBE,SINGO,WA102,DOLAN}}. Such a physical state would represent in production mode, regardless 
its further evolution to $q\bar{q}$, a state of two glueballs with negligible interaction energy~\cite{MORNI,MATHI}, as it can be expected for colorless states. This two-glueball state is also predicted by lattice QCD calculations~\cite{MORNI}. 
Such a state does not necessarily exist, but if it exists then it would provide a signature of the Pomeron-glueball equivalence and automatically constrain the glueball mass. 
Moreover, in analogy to equivalent diagrams in Electroweak Theory and related observed effects, the individually quasi-resonant status of the two real Pomerons generating the observable final state may be suggestive of an enhanced resonating behaviour of this state~\cite{EEZWW}. 
It should also be noted that even a system of two virtual DPE Pomerons, given an available global invariant mass greater or equal to the threshold of relation~(\ref{2}), could materialise into a two-glueball final state, whose cross-section could be compared to other two-particle final states. 
This possibility might be tested at hadronic colliders under dedicated conditions. 

It can be concluded therefore that the spectrum of $0^{++}$ scalar states could include a glueball physics state (e.g. one of the$f_0$ candidates) and a resonant state with mass equal to twice the glueball rest mass.

\begin{figure}
\centering
\includegraphics[scale=0.4]{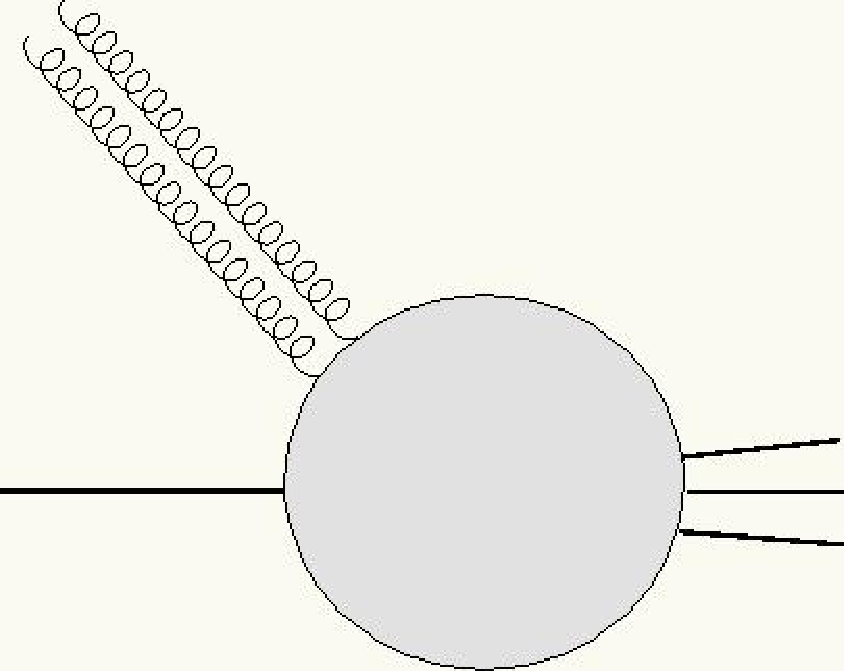}
\caption{\label{fig:epsart2} A symbolic representation of Pomeron-proton scattering diagram in either $s$ or $t$ channels.  
}
\label{spe}
\end{figure}

Following the same logic, the case of Single Diffraction (SD) can be also analysed. A single Pomeron, i.e.~a virtual glueball, may be exchanged between two scattering protons, leading to the diffractive dissociation of one of the two protons (Fig.~\ref{spe}).

The Single Diffraction cross-section is described by the following expression~\cite{ABRAM}:
\begin{equation}\label{sde}
\sigma_{\cal{SD}}(s)=\int F^{\cal{P}}(\xi, \hat{q}^2) \;\hat{\sigma}(\hat{q}^2)\; d\xi  
\end{equation}

where $F^{\cal{P}}(\xi, \hat{q}^2)$ is the diffractive structure function of the proton and $\hat{\sigma}(\hat{q}^2)$ is
the Pomeron-proton cross-section to produce a diffractive mass system.

In a fixed target experiment, it can be considered the special case of a projectile proton emitting a Pomeron with a negative four-momentum squared. If, as discussed above, we consider the dual diagram where the $s$-channel is exchanged with the $t$-channel $(s \leftrightarrow t) $ the virtual Pomeron  becomes a real glueball (from the $p\bar{p}$ annihilation) interacting with a proton at rest. If the glueball is such that it is close to its mass shell threshold, it follows that $P_{\cal{P}}^2 \cong M_{\cal{P}}^2 \cong M_{gg}^2$ with $M_{gg}^2$ being the glueball mass squared. The target proton at rest absorbing such a real Pomeron will produce a mass system $M_x$. The production of physics states will occur respecting the condition given by the relation (\ref{4}):
\begin{equation}\label{4}
( P_{\cal{P}} + P_{proton} )^2 \geq (M_{gg} + M_p)^2 ,
\end{equation}
where  $P_{\cal{P}}$ and $P_{proton}$ are the four-momenta of the real Pomeron and the target proton,  $M_{gg}$ is the glueball rest mass and  $M_{p}$  is the proton mass. The relation~(\ref{4}) allows a production threshold for a positively charged baryonic state with a mass equal to the glueball rest mass plus the proton rest mass\footnote{The same considerations as for DPE on the four-momenta scalar product are applicable and the conditions chosen for such a fixed-target experiment automatically satisfy the requirements.}. Such a physical state, produced by two combining colorless states, would represent either a deuterium-like bound state of a proton and a glueball with negligible binding energy, or a merged state of a proton and a glueball with negligible interaction energy between them, which could also be associated to the effect of color transparency~\cite{MUELL}. 
Also such a state, kinematically allowed, does not necessarily exist dynamically, but if it exists then it would provide an additional signature of the Pomeron-glueball equivalence and automatically constrain further the glueball mass.

Therefore it may be concluded that the spectrum of particles with baryon number $B=1$ could contain a positively charged resonant state with a mass equal to the sum of the glueball and proton rest masses. 

Finally, the case where a $2^{++}$ and a $0^{++}$ on-shell Pomerons interact to create a $2^{++}$ physics state (for Landau-Yang's theorem) can be analysed. These conditions, following the same logic as before, can foresee a production threshold for a $2^{++}$ mesonic state via the relation~(\ref{rel_0_2_0_2}):
\begin{equation}\label{rel_0_2_0_2}
( P_{{\cal{P}},0^{++}} + P_{{\cal{P}},2^{++}} )^2 \geq (M_{gg} + M_{gg,2^{++}})^2,
\end{equation}
where $M_{gg,2^{++}}$ is the mass of the $2^{++}$ glueball. Consequently, a $2^{++}$ mesonic state could exist with a mass equal to the sum of the scalar and tensorial glueballs' masses. Such a combination of states do not necessarily exist, but if they exist then they would provide a signature for the mass of the $2^{++}$ glueball. 

It can be concluded therefore that the spectrum of $2^{++}$ mesonic states could include two states such that their mass difference equals the scalar glueball mass.

\section{Experimental observations}
Searching in the experimentally measured spectrum of $0^{++}$ scalar states and checking confirmed states with uncertainties on the central mass value not greater than $10\UMeV$~\cite{PDG12}, it can be observed that the pair $f_0(1710)$ and $\chi_{c0}(1P)$ exists with a mass ratio of 2 within the experimental precision. No other pairs of $0^{++}$ states obeying to the same constraints exist in the $0 - 10\UGeV$ range.

The $\chi_{c0}$ has a mass of $3414.75 \pm 0.31\UMeV$~\cite{PDG12} and, therefore, one half of its mass equals $1707.4 \UMeV$ with a precision of about $150\UkeV$. The relation between the $c\bar{c}$ structure of the $\chi_{c0}$ and its gluonic features has been discussed in the literature~\cite{LUCAT}. It should also be noted that the evolution of glueball pairs into charmonium states is made natural by the relation between the $1.7 \UGeV$ mass scale and the charm vacuum extraction threshold given its Compton wavelength.

Furthermore, if such an evolution occurred by conserving the
mass before the decay, it would not be necessary to assume that
glueballs (single and pairs of) states need to be in excess to the
$SU(3)_f$ multiplets”.

The $f_0(1710)$ has a mass of $1701\pm 5 ^{+9}_{-2}\UMeV$ or of $1707 \pm 10\UMeV$ according to the two most precise experimental measurements published in the literature~\cite{ZEUS,AUGU}. Both of them are consistent with one half of the $\chi_{c0}$ mass within one standard deviation. For the $f_0(1710)$ mass the Particle Data Book 2012~\cite{PDG12} shows an ideogram-based most probable value of about $1704 \UMeV$ and a weighted average value of $1720\UMeV$, with an uncertainty of  $\pm 6\UMeV$ (the $\chi^2$ contributions~\cite{PDG12} show that the incompatibility between them is essentially due to a measurement given at $1765\UMeV$~\cite{BES2}).

From a theoretical perspective, QCD lattice calculations~\cite{YCHEN,MORNI,MATHI} foresee the glueball $0^{++}$ ground state at $1710\pm 50\pm 80 \UMeV$~\cite{YCHEN}. Given tolerances include also the model/calibration uncertainties introduced when moving from the pure gauge theory (quenched approximation) to QCD taking into account quarks and the instability of glueballs~\cite{LUSCH,YCHEN,MORNI,GREGO,MATHI}. The $f_0(1710)$ is the only currently observed $0^{++}$ state strictly fitting these limits. 
Moreover, coupling to the $s$-quark in the decay mode has been confirmed, by chiral symmetry suppression considerations~\cite{CHANO,WEING}, to be a key property of glueballs consistently with the observed branching ratios of the $f_0(1710)$. 

Contrarily to previous studies~\cite{QZHAO,DOLAN,PDG06,WOCHS}, also recent phenomenological works~\cite{LEEWE,HAIYA,DEWIT,CREDE} have developed glueball-mesons mixing scenarios where the $f_0(1710)$ results as a gluonic state at $\approx100\%$ purity level 
and where the $u\bar{u}$/$d\bar{d}$, $s\bar{s}$, $gg$ mass hierarchy is more consistent with current QCD predictions. Moreover, it has been argued~\cite{HASHI} within the holographic-QCD framework that mixing of mass eigenstates between glueballs and mesons is excluded. The mixing occurs and is observable only in the final state decay modes. This argument would open the perspective of purely measuring glueballs at the LHC focusing on the DPE production via initial states determined by the protons measurement. Experimental evidence of a peculiar behaviour in the 
differential cross-section of a diffractive production channel at $\sqrt{-t} \approx 1.7\UGeV$ has been already observed~\cite{COTAN,ANCIA} at TJNAF. 

In summary, the glueball mass $M_{gg}$ can be estimated (with tolerances of $\sim 1\UMeV$) as:
\begin{equation}\label{5}
M_{gg} \cong 1707.4\UMeV \cong \frac{1}{2} M_{\chi_{c0}} \cong \frac{1}{2} \cdot 3414.75\UMeV\\ \cong M_{f_0} \cong 1701 \pm 5 ^{+9}_{-2} \UMeV \cong 1707 \pm 10\UMeV .
\end{equation}
Searching in the experimentally measured spectrum of baryons, it can be observed that a positively charged narrow $\Xi_{c}^{+}$ charmed state exists with a mass of $2645.9 ^{+0.5}_{-0.6}\UMeV$~\cite{PDG12}. The $\Xi_c^+$ mass is exactly equal to the sum of the proton mass and one glueball rest mass (one-half of the $\chi_{c0}$) within the extremely constraining precision given by their experimental values and accuracies.
 
In fact, the $\Xi_c^+$ mass can be computed (with tolerances of $\sim 100\UkeV$) as:
\begin{equation}\label{6}
M_{\Xi_c^+} \cong M_p + M_{gg} \cong 938.27\UMeV + 1707.4\UMeV \cong 2645.7 \UMeV \cong 2645.9 ^{+0.5}_{-0.6}\UMeV.
\end{equation}
The existence of hadronic states where a glueball co-exists within e.g. a proton with negligible binding energy is suggestive of several implications. The charm valence of the obtained $\Xi_c^+$ state indicates an intimate interaction between the proton and the glueball, as opposed to a deuterium-like binding. 
As a possible mechanism we may consider the glueball-proton interaction to take place as a two step process consisting in $u$ and $d$ quark states being, as a first step, excited to $u^*$ and $d^*$ states and, subsequently, transformed to $s$ and $c$ quark states respectively, i.e. $u^* \rightarrow s$ and $d^* \rightarrow c$, by a $W$ internal exchange giving the $\Xi_c^+$ flavour content. 
These results are related to specific features of the strong interaction and may be related to color transparency~\cite{MUELL} within the nucleon when it interacts with a gluonic state with no relative momentum as opposed to the standard scattering case. 
The spin of the $\Xi_c^+$ has not been measured yet~\cite{PDG12}. Given the low relative momentum between the on-shell Pomeron and the proton, just allowed by the $f_0$(1710) width, and given a typical impact parameter, both the values $\tfrac {1}{2}$ or $\tfrac {3}{2}$ are conceivable for the $\Xi_c^+$ spin. 

Searching in the experimentally measured spectrum of $2^{++}$ states, it can be observed that the $\chi_{c2}(2P)$ resonance exists with a mass of 3927.2 $\pm$ 2.6$\UMeV$. This is such that, subtracting to it the scalar glueball mass $M_{gg}$ (1707.4$\UMeV$), it would foresee a $2^{++}$ glueball candidate mass $M_{gg,2^{++}}$ in the range 2217.2 - 2222.4$\UMeV$. These values are compatible with the most recent determination of the $f_J$(2220) $0^+(2^{++})$ mass, which equals $2223.9 \pm 2.5\UMeV$~\cite{VLADI}. For the same resonance, different experiments~\cite{PDG12} found mass values of, respectively, $2232 \pm 7 \pm 7\UMeV$, $2220 \pm 10\UMeV$, $2230 \pm 20\UMeV$, $2209^{+17}_{-15} \pm 10\UMeV$, $2235 \pm 4 \pm 5\UMeV$. 
Lattice QCD calculations foresee the $2^{++}$ glueball mass at a value higher than the $f_J(2220)$ by slightly more than one sigma of their uncertainty~\cite{YCHEN}.  
The $f_J(2220)$ is currently considered in the literature\cite{ZHENG,ANISO} a well fitting candidate for being the $2^{++}$ glueball.

In summary:
\begin{equation}\label{7_2220}
M_{gg} + M_{gg,2^{++}} \cong 1707.4\UMeV + 2223.9 \pm 2.5\UMeV \cong 3931.3 \pm 2.5\UMeV \cong 3927.2 \pm 2.6\UMeV \cong M_{\chi_{c2}}
\end{equation}

The coincidence of glueball-induced states with $SU$ foreseen quark states (which can be produced also otherwise) can be regarded as natural given the continue transformation between gluons and quarks within the hadron as an isolated system (which conserves the mass). The glueball induced channel can be seen as a channel indicating the mass value.

Finally, it can be supposed that by combining two tensor glueballs or a tensor glueball with a proton one could expect a $0^{++}$, $2^{++}$ or $4^{++}$ meson state in the range of $4440\UMeV$ and a baryon in the range of $3160\UMeV$. 

\section{Conclusions} 
 
Kinematic constraints, long-range strong interaction dynamics and $s$ -- $t$ duality have been used to search for states of pure gluonic contents. Consistently with the formalization of Pomerons as virtual glueballs and with the pQCD lattice calculations, it can be concluded that the scalar $f_0(1710)$ is compatible to be a glueball of mass $M_{gg} \cong 1707.4 \UMeV$ and the $\chi_{c0}$ is compatible to be  a $c\bar{c}$ scalar state produced from a pair of glueballs. The same method has been used to search for bound states of glueballs and hadrons. It is found that the charmed baryon $\Xi_c^+(2645)$ can be a state precisely produced by a proton and a glueball merging with negligible binding energy. The same method has also been used to constrain the mass of the tensorial glueball, finding results compatible with the $f_J(2220)$ being the $2^{++}$ glueball candidate.



\end{document}